%% file: main.tex
\documentclass[runningheads]{llncs}

\usepackage[T1]{fontenc}
\usepackage{graphicx}
\usepackage{xspace}

\usepackage{booktabs}
\usepackage{subcaption}
\usepackage{threeparttable}
\usepackage{todonotes}
\usepackage{amsmath}
\usepackage{algorithm}
\usepackage{algpseudocode}
\usepackage[hidelinks]{hyperref}
\usepackage[nameinlink]{cleveref} %
\usepackage{tikz}  

\DeclareRobustCommand{\circled}[1]{%
  \tikz[baseline=(char.base)]{%
    \node[draw,circle,inner sep=1.1pt,line width=0.5pt] (char)
      {\sffamily\bfseries\footnotesize #1};%
  }%
}
\usepackage{tcolorbox}

\newcommand{\tool}{\textsc{GreenAFL}\xspace}
\newcommand{\aflpp}{\textsc{AFL++}\xspace}

\usepackage{listings}
\usepackage{xcolor}
\usepackage{caption}
\definecolor{mGreen}{rgb}{0,0.6,0}
\definecolor{mGray}{rgb}{0.5,0.5,0.5}
\definecolor{mPurple}{rgb}{0.58,0,0.82}
\definecolor{backgroundColour}{rgb}{0.99,0.99,0.97}
\lstdefinestyle{CStyle}{
    backgroundcolor=\color{backgroundColour},   
    commentstyle=\color{mGreen},
    keywordstyle=\color{magenta},
    numberstyle=\tiny\color{mGray},
    stringstyle=\color{mPurple},
    basicstyle=\tt\fontsize{7}{8}\selectfont,
    breakatwhitespace=false,         
    breaklines=true,                 
    captionpos=b,
    keepspaces=true,                 
    numbers=left,
    stepnumber=1,
    firstnumber=1,
    numbersep=3pt,                  
    showspaces=false,                
    showstringspaces=false,
    showtabs=false,                  
    tabsize=2,
    aboveskip=-0.01pt,
    belowskip=-0.01pt,
    language=C
}

\makeatletter
\renewcommand\section{%
  \@startsection{section}{1}{\z@}%
    {0.8ex \@plus .2ex \@minus .2ex}%
    {0.4ex}%
    {\normalfont\Large\bfseries}%
}

\renewcommand\subsection{%
  \@startsection{subsection}{2}{\z@}%
    {0.8ex \@plus .2ex \@minus .2ex}%
    {0.4ex}%
    {\normalfont\large\bfseries}%
}

\renewcommand\subsubsection{%
  \@startsection{subsubsection}{3}{\z@}%
    {0.8ex \@plus .2ex \@minus .2ex}%
    {0.4ex}%
    {\normalfont\normalsize\bfseries}%
}
\makeatother

\makeatletter
\renewenvironment{abstract}{%
    \if@twocolumn
      \section*{\abstractname}%
    \else
      \small
      \begin{center}%
        {\bfseries \abstractname\vspace{-.5em}\vspace{\z@}}%
      \end{center}%
      \quotation
    \fi}
    {\if@twocolumn\else\endquotation\fi}
\makeatother

\usepackage{booktabs}
\usepackage{siunitx}
\usepackage{multirow}
\usepackage{makecell} %

\begin{document}

\title{Fuzz Smarter, Not Harder: Towards Greener Fuzzing with \tool{}}
\author{Ayse Irmak Ercevik\inst{1}\orcidID{0009-0000-0974-2527} \and
Aidan Dakhama\inst{1}\orcidID{0009-0002-7318-7964} \and
Melane Navaratnarajah\inst{1}\orcidID{0009-0001-8987-6134} \and
Yazhuo Cao\inst{1}\orcidID{0009-0002-1201-9908} \and
Leo Fernandes\inst{2}\orcidID{0000-0001-9090-2232}}
\authorrunning{Ayse Irmak Ercevik et al.}
\institute{King’s College London, London, UK.
\email{\{ayse.ercevik, aidan.dakhama, melane.navaratnarajah, yazhuo.cao\}@kcl.ac.uk} \and
The Federal Institute of Education, Science, and Technology of Alagoas, Brazil.
\email{leonardo.fernandes@ifal.edu.br}}

\maketitle              %
\begin{abstract}
Fuzzing has become a key search-based technique for software testing, but continuous fuzzing campaigns consume substantial computational resources and generate significant carbon footprints. Existing grey-box fuzzing approaches like AFL++ focus primarily on coverage maximisation, without considering the energy costs of exploring different execution paths. 
This paper presents \tool, an energy-aware framework that incorporates power consumption into the fuzzing heuristics to reduce the environmental impact of automated testing whilst maintaining coverage.
\tool{} introduces two key modifications to traditional fuzzing workflows: energy-aware corpus minimisation considering power consumption when reducing initial corpora, and energy-guided heuristics that direct mutation towards high-coverage, low-energy inputs. We conduct an ablation study comparing vanilla AFL++, energy-based corpus minimisation, and energy-based heuristics to evaluate the individual contributions of each component.
Our evaluation shows up to 7.4\% lower energy usage and 7.1\% lower throughput while maintaining or improving coverage, with best-case coverage gains of 2.6\%.

\end{abstract}
\input{sections/introduction}
\input{sections/tool}

\input{sections/results}
\input{sections/conclusions}

\vspace{-0.1cm}
\begin{credits}
\subsubsection{\ackname} We thank CloudLab \cite{duplyakin2019design} for providing the platform and infrastructure that enabled our experiments.

\end{credits}
\bibliographystyle{splncs04}
\bibliography{main}
\end{document}

%% file: sections/introduction.tex
\section{Introduction}

Fuzzing has emerged as a powerful greedy search-based technique for discovering bugs in complex software systems, serving as a common search-based technique in industrial testing pipelines; however, it comes with significant computational and energy costs. Systems such as OSS-Fuzz~\cite{ossfuzz} run continuous fuzzing campaigns using substantial resources, consequently producing large carbon footprints. Similarly, fuzzing is increasingly applied to large-scale and energy-intensive systems such as operating systems~\cite{bursey2024syzretrospector,shi2019industry}, quantum system simulators~\cite{blackwell2024fuzzing}, and system simulators~\cite{dakhama2023searchgem5,dakhama2025enhancing,even2025search+}. As automated testing scales, it becomes increasingly important to manage the energy usage of such systems.
Grey-box fuzzing techniques such as \aflpp use coverage feedback to guide input mutation towards more effective test cases. Whilst this coverage-guided approach improves the efficiency of discovering new execution paths, it fails to consider the energy consumption associated with exploring these paths. Consequently, running fuzzing campaigns remains an energy-intensive process.
Much of the existing work on reducing the emissions of fuzzing focuses on stopping criteria~\cite{lipp2023green}, or in the case of \texttt{GreenBench}, improving the emissions of the benchmarks used to evaluate fuzzers~\cite{ounjai2023green}. Yue et al.’s EcoFuzz reduces redundant test case generation through optimised scheduling with an Adversarial Multi-Armed Bandit model, achieving higher coverage from fewer executions~\cite{yue2020ecofuzz}. Similarly, Lyu et al.’s SLIME introduces program-sensitive energy allocation to adaptively distribute fuzzing effort across seeds rather than measuring actual power use~\cite{lyu2022slime}. Unlike our approach, neither EcoFuzz nor SLIME measure actual energy usage, but instead optimise seed scheduling to indirectly reduce waste.

We propose \tool, a modification to \texttt{afl-cmin} that considers the system energy usage of each input when minimising the initial corpus. Further, we extend the heuristics used by \aflpp to guide fuzzing towards inputs that achieve high coverage whilst maintaining low CPU and memory energy consumption. To the best of our knowledge, this is the first work that directly incorporates energy usage into fuzzing and minimisation heuristics.
In our evaluation, energy-aware corpus minimisation consistently delivered the best balance of reduced energy (up to 7.4\%) and maintained or improved coverage (up to 2.6\%), while energy-guided fuzzing heuristics showed promising benefits but require further refinement. Together, these results demonstrate that direct integration of energy awareness into fuzzing workflows is both feasible and impactful. Beyond sustainability, reducing energy translates directly into lower infrastructure costs, making greener fuzzing attractive for industrial-scale deployments.

\vspace{0.05cm}
\noindent\textit{Availability. }
\tool{}, 
and documentation demonstrating its generalisability, reproducibility materials, and experimental results, are available at \cite{zenodo:greenafl}.

%% file: sections/tool.tex
\section{\tool}
\tool extends the AFL++ fuzzing framework by integrating energy awareness into corpora minimisation (\texttt{green-cmin}) and energy-aware fuzzing heuristics (\texttt{green-afl}). Our approach leverages Intel's hardware-based power monitoring via \texttt{cppjoules}~\cite{chattaraj2024cppjoules}. Our implementation is generic and modular, utilising the \texttt{LD\_PRELOAD} library that wraps the system under test to record energy usage to ensure portability. While our implementation is Intel-specific, the preload-based mechanism can be adapted to other architectures (e.g.\ AMD or ARM CPUs) and extended to NVIDIA GPUs, where \texttt{cppjoules} already provides support~\cite{chattaraj2024cppjoules}.

\subsection{Energy-Aware Corpus Minimisation}

AFL’s \texttt{cmin} reduces the initial corpus by discarding redundant inputs, keeping only those that contribute new coverage. \tool augments this process by also considering the energy consumption of each input. Specifically, we retain inputs that maximise coverage while minimising power cost -- balancing effectiveness with efficiency; this can reduce the emissions required for a test, as well as serve to reduce running costs, especially in the context of large scale long running campaigns. This ensures that the subsequent fuzzing loop operates on inputs that are both coverage-rich and energy-efficient. \tool runs \verb|afl-showmap| once per seed to record which program edges the seed reaches, while measuring the total energy for that run. For each edge, we keep the seed that has the lowest energy cost. It is important to note that this measurement reflects the energy consumed during the execution of each seed by the fuzzer, rather than the long-term cost of the resulting test archive.

\subsection{Energy-Guided Fuzzing Loop}
\label{sec:FuzzingLoop}

In traditional \aflpp, the fuzzing loop schedules mutant generation according to new coverage. \tool extends this loop with energy-aware heuristics that modify the allocation of mutation cycles.
These heuristics aim to bias the search towards executions with higher ``coverage-per-watt''.

\begin{figure}[!t]
  \centering
  \vspace{-0.4cm}
  \includegraphics[width=0.85\linewidth]{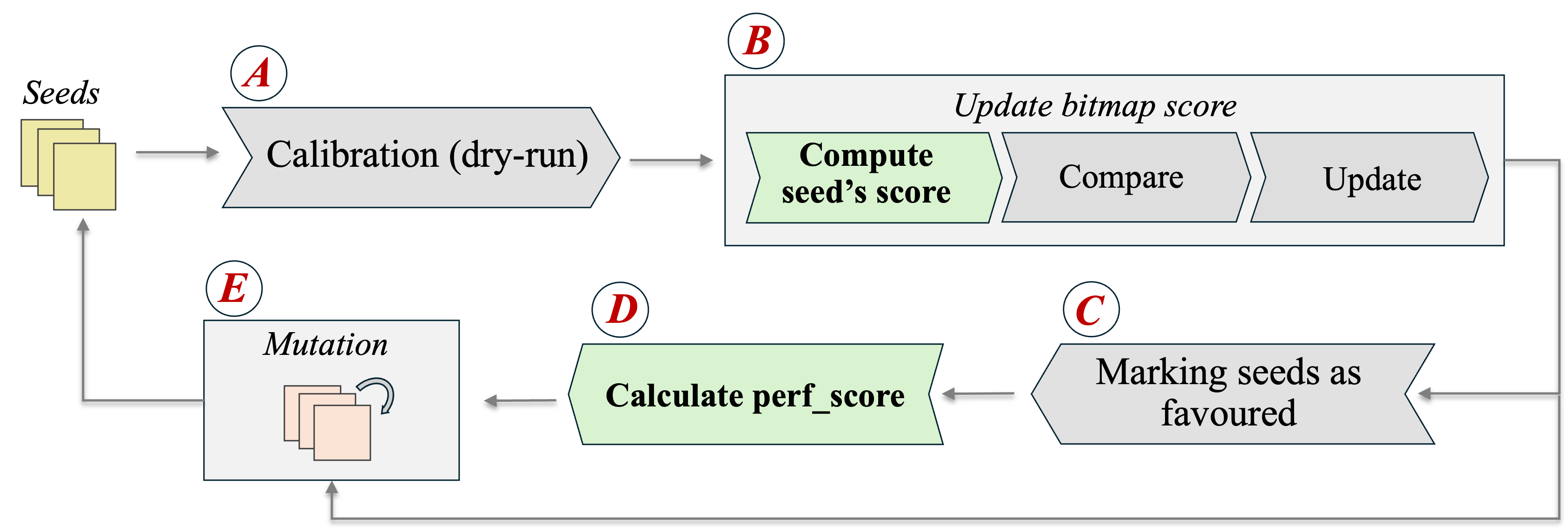}
  \vspace{-0.1cm}
  \caption{Overview of \tool’s energy-guided fuzzing loop. The {\textbf{green boxes with bold text}} highlight where our energy-aware heuristics are applied, (\circled{\textit{B}}, energy-aware score computation) and (\circled{\textit{D}}, airtime scheduling).} 
  \vspace{-0.2 cm}
  \label{fig:greenafl-diagram}
\end{figure}

\subsubsection{Airtime Scheduling via Energy-Scaled Performance Score} 

In the \aflpp fuzzing loop (\cref{fig:greenafl-diagram}, \circled{\textit{D}}) each seed has a \verb|perf_score| computed which determines how much fuzzing effort is allocated and represents a combination of factors -- such as execution speed and input size -- which is used to determine the number of mutation cycles.
We modify \verb|perf_score| to be energy‑aware by applying a scale factor inversely proportional to the energy used during the execution of the seed. The multipliers are derived from each seed’s CPU and memory energy relative to the campaign’s global minimum and maximum energy cost, which we map to the range of $5\times$ to $\frac{1}{5}\times$. This prioritises low energy seeds, allowing for more mutation cycles (\cref{fig:greenafl-diagram}, \circled{\textit{E}}).

\subsubsection{Energy-Aware Favoured Seed Selection} 

\aflpp assigns a single champion seed to each edge transition in the coverage bitmap (\cref{fig:greenafl-diagram}, \circled{\textit{B}}). 
Whenever a seed is executed and hits an instrumented edge, its score is compared against the current champion’s score. If the new seed has a lower score, it replaces the current champion for that edge. We make this selection energy-aware by applying an energy-based scale factor in the range of $\frac{4}{5}\times$ to $\frac{5}{4}\times$. The multiplier is obtained by a logarithmic normalisation of the seed’s total energy cost relative to the campaign's minimum and maximum energy costs. This prioritises low-energy seeds and penalises high-energy seeds, marking low-energy seeds as favoured (\cref{fig:greenafl-diagram} stage \circled{\textit{C}}). Favoured seeds are then prioritised in the fuzzing schedule and selected more often for mutation.

\subsection{Experimental Setup}
We evaluate GreenAFL through an ablation study with two toggles: corpus minimisation (\texttt{green-cmin}, \texttt{afl-cmin}) and energy-guided fuzzing (\texttt{green-fuzz}, \texttt{afl-fuzz}). We run four configurations – neither, each individually, and both. The impact of each setting is measured in terms of total energy consumption, as reported by \texttt{perf stat}. The fuzzing targets come from OSS-Fuzz~\cite{ossfuzz}, specifically we explore \texttt{libpng}, \texttt{zlib}, and \texttt{jsoncpp} with the corpora and test harnesses provided by OSS-Fuzz. We run this evaluation over three repetitions across each configuration and each target, with 24 hour campaigns. We perform this evaluation on an Intel Xeon D-1548 CPU (2.0 GHz, 8 cores), with 64 GB RAM, 8 GB swap, and running Ubuntu 22.04.5 LTS (x86\_64), ensuring only one campaign is run at a given time to minimise external effects on power usage.

%% file: sections/results.tex
\section{Results}
\begin{table*}[t]
  \centering
  \small
  \setlength{\tabcolsep}{3pt} 
  \renewcommand{\arraystretch}{1.4}
  \resizebox{\textwidth}{!}{%
  \begin{tabular}{llccc ccc ccc}
    \toprule
    \multicolumn{2}{c}{\textbf{Config}} &
    \multicolumn{3}{c}{\textbf{Throughput}} &
    \multicolumn{3}{c}{\textbf{Energy (kJ)}} &
    \multicolumn{3}{c}{\textbf{Coverage (\%)}} \\
    \cmidrule(lr){1-2}\cmidrule(lr){3-5}\cmidrule(lr){6-8}\cmidrule(lr){9-11}
    \textbf{cmin} & \textbf{fuzz} &
    \textit{libpng} & \textit{zlib} & \textit{jsoncpp} &
    \textit{libpng} & \textit{zlib} & \textit{jsoncpp} &
    \textit{libpng} & \textit{zlib} & \textit{jsoncpp} \\
    \midrule
    afl & afl &
    3135 $\pm$ 15.6 & 2845 $\pm$ 54 & 2001 $\pm$ 170 &
    2122 $\pm$ 1.74 & 2029 $\pm$ 7.12 & 2021 $\pm$ 6.31 &
    0.2 $\pm$ 0 & 50.1 $\pm$ 0.28 & 38.9 $\pm$ 0 \\
    green & afl &
    1110 $\pm$ 871 & 2947 $\pm$ 43.2 & 2133 $\pm$ 59.1 &
    \textbf{2120} $\pm$ 40.2 & 2153 $\pm$ 10.2 & \textbf{\textit{1872}} $\pm$ 4.73 &
    39.5 $\pm$ 0.01 & \textbf{\textit{51.4}} $\pm$ 0 & \textbf{39.1} $\pm$ 0 \\
    afl & green &
    2101 $\pm$ 11.9 & \textbf{1867} $\pm$ 91.9 & \textbf{1826} $\pm$ 37.4 &
    2167 $\pm$ 1.22 & 2211 $\pm$ 3.29 & 2211 $\pm$ 10.2 &
    0.2 $\pm$ 0 & 50.3 $\pm$ 0.55 & 38.9 $\pm$ 0 \\
    green & green &
    \textbf{\textit{969}} $\pm$ 200 & 1981 $\pm$ 28.4 & 1858 $\pm$ 2.65 &
    2421 $\pm$ 7.41 & \textbf{1971} $\pm$ 3.58 & 2252 $\pm$ 3.09 &
    \textbf{39.6} $\pm$ 0.01 & \textbf{\textit{51.4}} $\pm$ 0 & \textbf{39.1} $\pm$ 0 \\
    \bottomrule
  \end{tabular}
  } %
  \vspace{2pt}
  \caption{\footnotesize
    Results from 3 repetitions of fuzzing campaigns across all three targets, showing the mean $\pm$ variance. 
    \textbf{Energy} (as reported by \texttt{perf}) combines CPU and RAM package power. 
    \textbf{Throughput} represents executions per second. 
    \textbf{Bold} marks the best per target, \textbf{\textit{bold italics}} the best overall.
  }
  \label{tab:fuzzing_campaign}
  \vspace{-0.2cm}
\end{table*}

\textbf{RQ1:} \textit{To what extent can energy-aware corpus minimisation and fuzzing reduce energy use whilst maintaining or improving coverage across different targets?}

We compare the impact of our \texttt{green-fuzz} and \texttt{green-cmin} fuzzing campaigns by highlighting their distinctive contributions in Table~\ref{tab:fuzzing_campaign}. Incorporating \texttt{green-cmin} was found to consistently maintain or improve coverage across all targets whilst having the lowest energy usage. For example, \texttt{jsoncpp}, \texttt{green-cmin} achieved 39.1\% coverage compared to the baseline 38.9\% coverage while consuming 1872kJ (vs 2021kJ), resulting in a statistically significant reduction in energy consumption (Welch’s t-test, $p < 10^{-4}$). \texttt{zlib} achieved the highest coverage (51.4\%), which is also statistically significant compared to the baseline (Welch’s t-test, $p < 0.02$), and 3\% less energy than the baseline with both modifications. However, \texttt{green-fuzz} produces coverage that is often comparable to the baseline but can sometimes increase the overall energy usage. In all cases, the best performing configuration included at least one of our modifications; showing that prioritising inputs by energy usage can also result in higher fuzzing performance and yield tangible savings in energy. Additionally, the configurations that explicitly resulted in the lowest energy consumption often reached the highest coverage.

\texttt{green-fuzz}, reduces the execution throughput (execs/s) considerably. For \texttt{libpng} there is a 69.2\% reduction in executions from the baseline, yet achieving better coverage. This shows that less effort is needed for comparable results to the baseline coverage results.
A deeper analysis is required to determine whether \texttt{green-fuzz}'s higher total energy stems from increased energy per execution, overhead in measurement, or the heuristic calculations. However, the fact that \texttt{green-fuzz} reaches comparable results with fewer executions suggests potential efficiency gains if this issue can be resolved.

\texttt{libpng} was found to have an unexpected anomaly in the configurations that used the standard corpus minimisation, where it produced an extremely low coverage -- this was found to be a consequence of the over-aggressive minimisation of \texttt{afl-cmin}, reducing the initial corpus to 1 seed.

\vspace{0.1cm}
\textbf{(RQ2)} \textit{To what extent can energy-aware corpus minimisation and energy-guided heuristics improve the rate of convergence?}

\begin{figure*}[t!]
    \centering
    \vspace{-0.3cm}
    \begin{subfigure}[b]{0.9\textwidth}
    \centering
    \includegraphics[width=\textwidth]{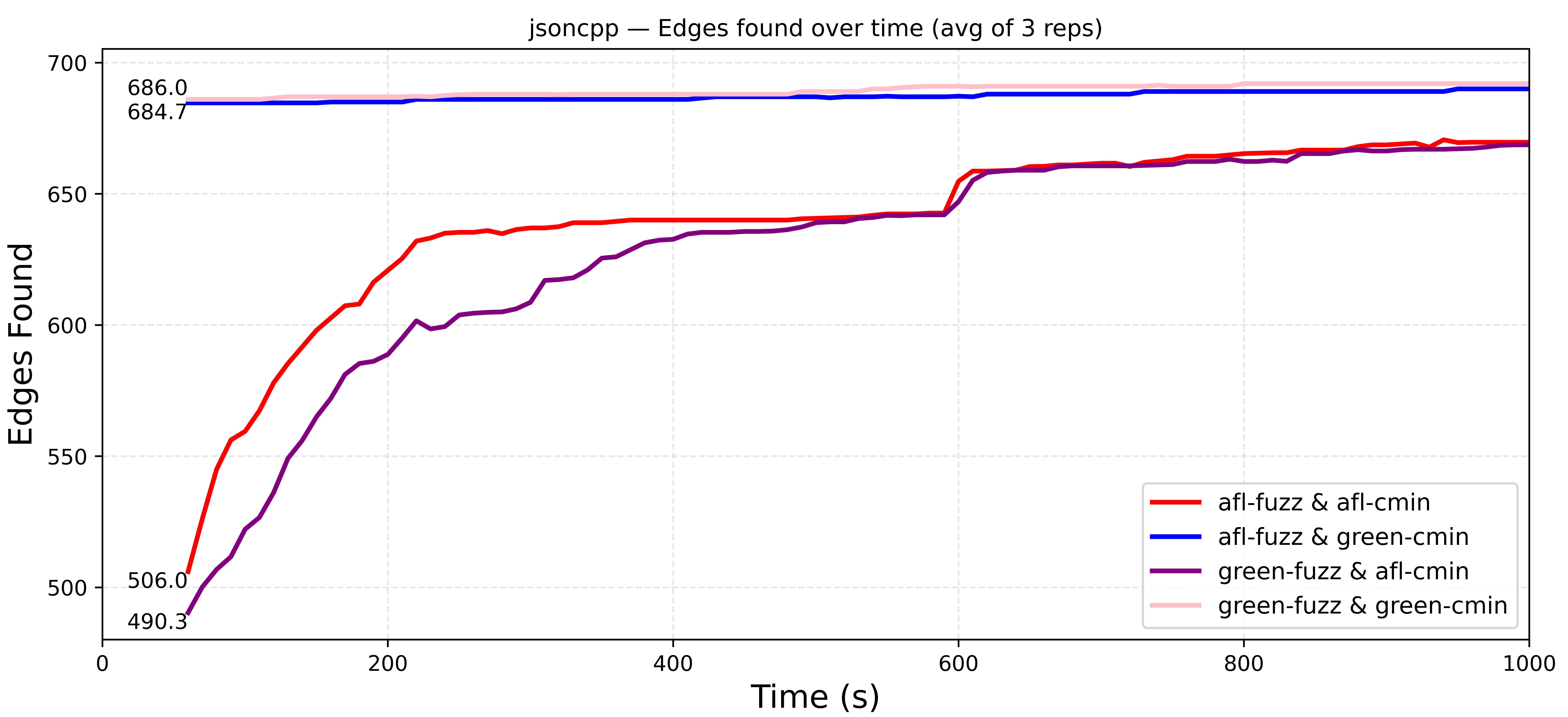}
    \end{subfigure}
    \hfill
    \vspace{-0.1cm}
    \caption{Edges found over time for \texttt{jsoncpp}. Each curve shows the mean across three repetition for a given configuration.}
    \vspace{-0.2 cm}
    \label{fig:edges_found}
\end{figure*}

Figure~\ref{fig:edges_found} highlights that runs with \texttt{green-cmin} have an early advantage; the energy-aware seeds find unique edges sooner than \texttt{afl-cmin}. For instance, \texttt{jsoncpp}'s first 680 unique edges were found within 60s, whereas this took more than 1200 seconds with \texttt{afl-cmin}. This pattern is present across all configurations and targets that include \texttt{green-cmin}, demonstrating the effectiveness of this modification. During the actual fuzzing for \texttt{green-cmin}, the coverage/edges only slowly increase, likely due to being closer to saturation of the targets. However, after the 24-hour runs, \texttt{afl-cmin} is still not surpassed by the baseline.
Although we do not have the time series for energy consumption, we can use throughput as a proxy. Inspection using throughput over time reveals that \texttt{green-fuzz} consistently reduces the execution throughput (execs/s) throughout the entire run, indicating that \texttt{green-fuzz} likely uses less energy in the actual execution of seeds despite the higher overall energy usage.

%% file: sections/conclusions.tex
\section{Discussion \& Conclusion}

In this work, we introduce \tool, an energy-aware extension of \aflpp that integrates energy costs into corpus minimisation and fuzzing heuristics. Our evaluation across a subset of industry standard OSS-Fuzz benchmarks demonstrates that integrating energy-guided heuristics into corpus minimisation \newline(\texttt{green-cmin}) is an effective method to increase coverage and reduce energy usage. \texttt{green-cmin} tends to preserve cheaper, coverage-rich seeds, giving an early advantage and preventing over-pruning.

While we did not find the energy-aware heuristic (\texttt{green-fuzz}) to reduce the energy cost of fuzzing, it showed promise in maintaining comparable coverage while executing fewer seeds. This shows that the bottleneck to lower energy usage is likely in the implementation and parameters of our heuristic. Further, we found a high fluctuation in the rate of execution when using \texttt{green-fuzz}, showing that the relationship between seed energy usage of mutants is not strictly linear. As such, future work includes tuning these parameters to a general range, exploring dynamically adjusting these parameters during the fuzzing process, and optimising the efficiency of the heuristic measurements and implementation.

Overall, our results show that energy-based heuristics can improve coverage-per-watt in fuzzing and help reduce the environmental impact of large-scale campaigns. Given the widespread use of fuzzing in industry (e.g., OSS-Fuzz~\cite{ossfuzz}), these results indicate that energy-aware extensions could offer immediate relevance for practical continuous testing pipelines, reducing the environmental footprint of large-scale fuzzing while improving efficiency.